\documentclass[prd,aps,preprint,tightenlines,superscriptaddress]{revtex4}
\usepackage{epsfig}
\usepackage{amsmath}
\begin{document}


\title{Polarized Gluon Distribution $\Delta g(x)$ in the Proton}
\author{Panying Chen}
\affiliation{Department of Physics, University of Maryland, College
Park, Maryland 20742, USA}
\author{Xiangdong Ji}
\affiliation{Department of Physics, University of Maryland, College
Park, Maryland 20742, USA} \affiliation{Center for High-Energy
Physics and Institute of Theoretical Physics, \\ Peking University
Beijing, 100080, P. R. China}

\date{\today}
\vspace{0.5in}
\begin{abstract}

We study the polarized gluon distribution $\Delta g(x)$ in a
longitudinally polarized proton. We argue that the distribution can
be calculated approximately from the quark color currents found in
simple quark models. The first result from the MIT bag model as well
as the non-relativistic quark model shows that $\Delta g(x)$ is
positive at all $x$. If this feature holds in QCD, it imposes a strong
constraint on phenomenological fits to experimental data. The total
gluon helicity $\Delta G$ from the bag model is about $0.3\hbar$ at
the scale of 1 GeV, considerably smaller than previous theoretical
expectations.

\end{abstract}

\maketitle

The polarized gluon distribution in the proton $\Delta g(x)$ has
been the focus of high-energy spin physics since the European
Muon Collaboration measurement of the quark helicity distributions
through polarized inclusive deep-inelastic scattering \cite{EMC}.
Gluons are known to play the key role in the proton's mass and
momentum \cite{massmomentum}, and it is expected that they
also play a
similar important role in the spin of the proton. This
expectation has motivated extensive experimental programs at DESY
(HERMES collaboration), CERN (COMPASS Collaboration), and polarized
RHIC. Much experimental progress has been made in probing the
polarized gluon distribution through leading hadron production,
charm production, and neutral pions and di-jets \cite{data}. High
statistics data expected from the polarized RHIC will provide a much
better picture on $\Delta g(x)$ in the near future \cite{futuredata}.

Two fundamental questions about the gluon polarization have attracted
the most attention: what it is total size $\Delta G=
\int^1_0dx\Delta g(x)$, and how does its sign vary with Feynman
momentum $x$. Both questions are difficult to answer in practice.
Since $\Delta G$ is not related to the matrix element of a local,
gauge-invariant operator, it cannot be calculated directly in
lattice QCD simulations, which have been the only non-perturbative
approach to solve the fundamental theory so far. Experimentally, it
is a challenge to measure $\Delta g(x)$ directly at a
fixed-Feynman $x$, with the exception of few channels at tree order
(e.g. direct photon production \cite{futuredata}). The phenomenological
distributions in the literature are obtained by ``educated"
parametrizations and fitting of the parameters to experimental data
\cite{fits}. The results depend on the functional forms
adopted, sensitive to assumptions such as if $\Delta g(x)$ is
allowed to change sign in $x$. Given the above situation, it is important
to investigate the possibility of calculating $\Delta g(x)$ in proton
models,
with the hope that some key features might be shared by QCD.

Gluons are known to play the dominant role in QCD. The gluons in the QCD
vacuum
are responsible for, among others, color confinement and chiral
symmetry breaking. Modeling these gluons is beyond the scope of this
study. On the other hand, the gluons in the unpolarized proton
contribute as much as 50\% of its momentum and mass
\cite{massmomentum}. These gluons are generated from the valence
quarks and affect strongly the dynamics of quarks in return.
Therefore they must be solved self-consistently with the motion of
the quarks, which is again outside the scope of this paper.

On the other hand, the polarized gluons---induced through quark
polarizations in a polarized proton---is a much smaller effect.
In fact, in QCD with a large number of colors $N_c$,
the polarized gluons are suppressed by $1/N_c^2$ related to those
in the QCD vacuum. As a consequence, the
spin-dependent gluon potential $A^\mu$ may be solved from the
chromodynamic Maxwell equation,
\begin{equation}
       D_\mu F^{\mu\nu} = J^{\nu} \ ,
\end{equation}
with some reasonable modeling of the spin-dependent quark color
current $J^\mu$. This situation is analogous to the small-$x$ gluons
which are calculable from the valence quark color charges in the
saturation region \cite{smallx}.

SU(6) quark models have had some reasonable successes in describing the
valence quark structure of the proton. For example, they give a reasonable
account of the magnetic moment of the proton. In
particular, the signs and magnitudes of the magnetic moments of the
up and down quarks are correlated postively with the total angular
momentum carried by them ($\mu_u = 3.6$, $\mu_d = -1.0$, vs. $J_u = 4/3$
and $J_d = -1/3$). What about the proton spin for which the naive
quark model prediction seems to have failed so badly? Well,
in the MIT bag model, although the spin is carried entirely by
quark angular momentum in the lowest-order wave function,
about 40\% comes from the orbital motion of the quarks \cite{jaffemanohar}.
Once the gluon contribution is taken into account by
the higher Fock states, the quark contribution must be renormalized
from these additional wave function components. If the gluons
contribute 50\%  of the proton spin, for example, the renormalized singlet
axial
charge in the bag model will be about $\Delta \Sigma =0.60/2 = 0.30$,
roughly consistent with the
current experimental data. Therefore, as a first estimate, the
polarized gluons may be calculated from the quark color
currents in the MIT bag model.

The total gluon polarization $\Delta G$ has been studied before in
quark models in Refs. \cite{jaffe, barone}. In Ref. \cite{jaffe},
the calculation is incomplete because only the interference diagram
has been included, and the contribution from a single quark
intermediate states has been ignored. The result is a negative
$\Delta G$. In Ref. \cite{barone}, the single quark contribution was
taken into account in the non-relativistic quark model and
$\Delta G$ is found to be positive after canceling the negative
interference contribution. As we shall see, a direct calculation of
$\Delta g(x)$ in non-local operator form produces actually a different
total $\Delta G$ that is consistent with parton physics.

The polarized gluon distribution $\Delta g(x)$ can be calculated as
a matrix element of the non-local operator \cite{Manohar:1990jx}
\begin{equation}
  \Delta g(x) = -\frac{i}{x}\int^\infty_{-\infty} \frac{d\lambda}{2\pi}
    e^{-i\lambda x} \langle P | F^{+\alpha}(\lambda n)W
    \tilde F^+_{~~\alpha}(0)|P\rangle \ ,
\end{equation}
where $|P\rangle$ is the proton state normalized covariantly, $n$ is
a light-like vector conjugating to an infinite momentum frame $P$.
$F^{\mu\nu}$ is the gluon field tensor and $W$ is a gauge link along
the direction $n$ connecting the two gluon field tensors, making the
whole operator gauge invariant. In this first calculation, we
neglect the effects of the nonlinear interactions, and then the gluons
fields behave as 8 independent Abelian fields. Under this
approximation, the gauge link can be ignored and an equivalent expression
is obtained by inserting a complete set of intermediate states between the
gluon field tensors,
\begin{eqnarray}
   \Delta g(x) &=& - i \int \frac{d^4k}{(2\pi)^4} \frac{1}{V_4}
     \epsilon^{\alpha\beta}_\perp (k^+ g^{\alpha\mu}
       g^{\beta\nu} - k^\alpha g^{+\mu} g^{\beta\nu}
        - k^\beta g^{+\nu} g^{\alpha\mu} )
        \delta(x - k\cdot n)  \nonumber \\
      && \times \sum_m \langle \tilde P |A_\mu^*(k)|m\rangle
        \langle m|A_\nu(k) |\tilde P\rangle (V_3\cdot 2P^+) \ ,
\end{eqnarray}
where $m$ sums over all possible intermediate states, and $V_3$ and
$V_4$ are the space and space-time volumes, respectively. The rescaled state
$|\tilde P\rangle$ is normalized to 1.

\begin{figure}[hbt]
\begin{center}
\includegraphics[width=6cm]{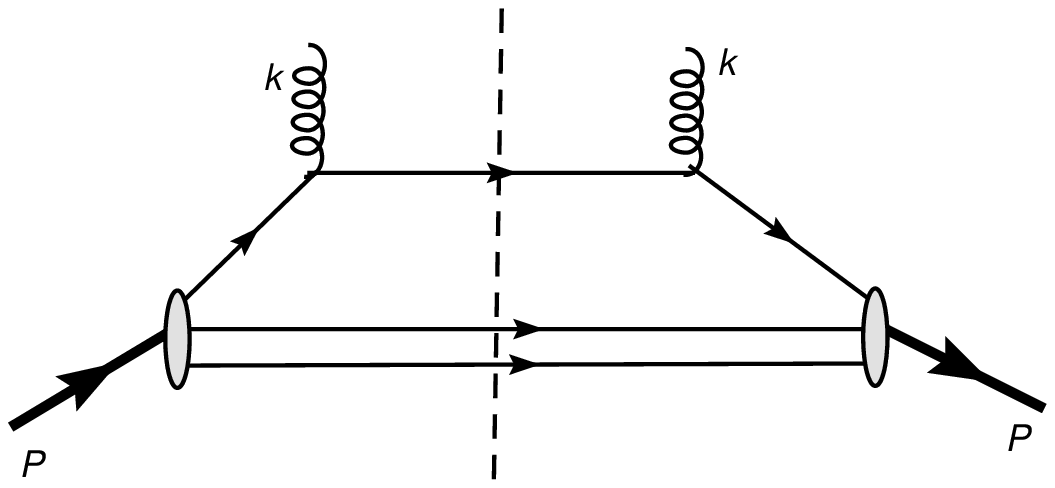}
\includegraphics[width=6cm]{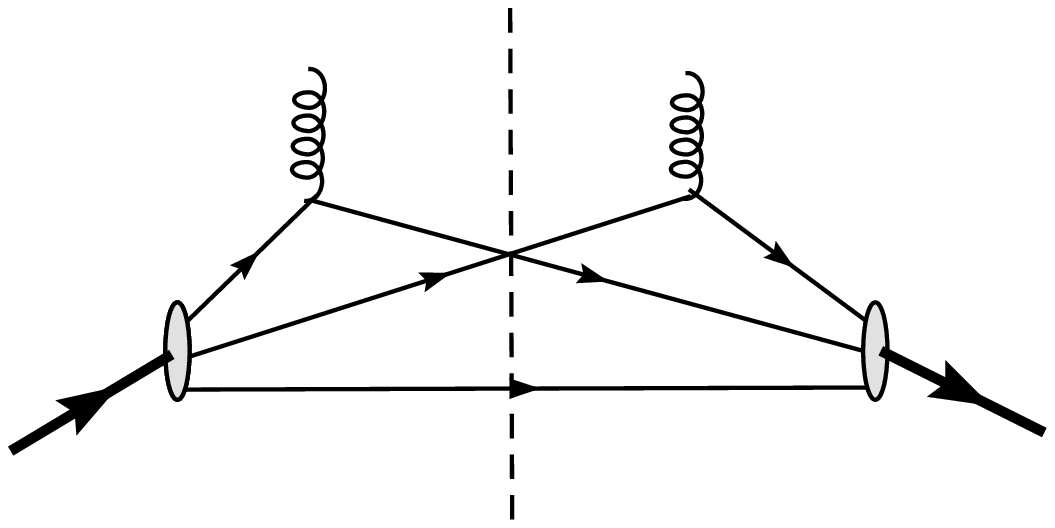}
\end{center}
\caption{One-body and two-body contributions to the matrix element
of the polarized gluon operator in the quark models of the proton. }
\label{Feynman-dg2}
\end{figure}

In the quark models, the above matrix element receives contributions
from one-body and two-body terms, shown schematically in Fig. 1. It is easy
to
show that the one-body term with the same quark intermediate state
as the initial one cancels the two-body contribution. This
cancellation is due to the color structure of the states as well as
the spin property of the operator. Therefore, the net contribution
arises from one-body term with excited intermediate quark states. The matrix
element in the single quark state is
\begin{equation}
   \langle m|A_\nu(k)|\tilde P\rangle
    = 2\pi \delta\big(k^0-(\epsilon_f - \epsilon_i)\big)
      \frac{-i}{k^2} (-igt^a) \langle m| j_\nu(k)|\tilde P\rangle \ ,
\end{equation}
where the $\delta$-function comes from the energy conservation and $j_\nu$
is the color current. For simplicity we have used the
free-space gluon propagator.

The sum over all intermediate quark states produces a divergent
result. This divergence is the usual ultra-violet divergence in
field theory and must be regulated by cut-offs. In our case, the
cut-off may be taken as the excitation energy of the intermediate
states.

Shown on the left panel in Fig. 2 is the MIT bag result for
$\Delta g(x)$ with different intermediate state cut-offs. The bag
radius $R$ is chosen to be 1.18 fm by fitting the nucleon charge {\it{r.m.s.}} radius. The dotted curve corresponds
to the $s$-wave contribution ($\kappa=-1$) with zero
and one node wave functions included.
The dash-dotted curve includes in addition the $p$ wave contribution
($\kappa =-2, 1$); the dashed curve contains the $d$ wave
contribution ($\kappa=-3,2$); and finally the solid curve includes
up to the $f$ wave contribution ($\kappa=-4,3$). Two features of the
result are immediately obvious. First, $\Delta g(x)$ is positive
everywhere as a function of $x$, and vanishes quickly as
$x\rightarrow 1$. Second, as more intermediate states are included,
$\Delta g$ gets uniformly larger. In fact, for higher intermediate
states, the result shall change with the cut-off following the
perturbative QCD evolution. Therefore, $\Delta g(x)$ at different
scales can be obtained approximately by limiting the excitation
energy of the intermediate quarks. The solid-line result roughly
corresponds to a cut-off at 1 GeV. Of course, the present cut-off
scheme is different from that of the perturbative dimensional
regularization and minimal subtraction. The difference can in
principle be calculated in perturbation theory.

\begin{figure}[hbt]
\begin{center}
\includegraphics[width=8cm]{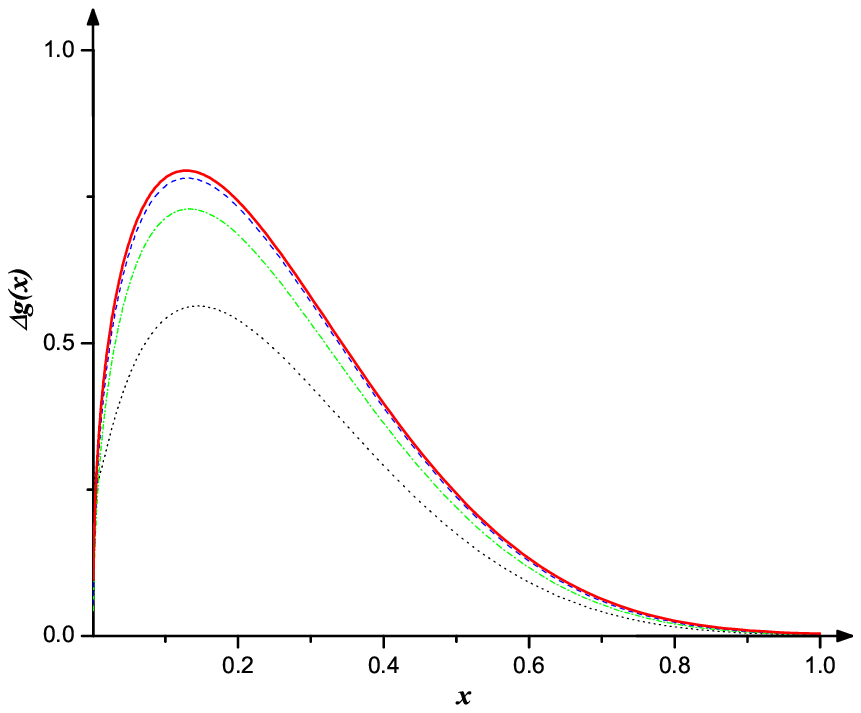}
\includegraphics[width=8cm]{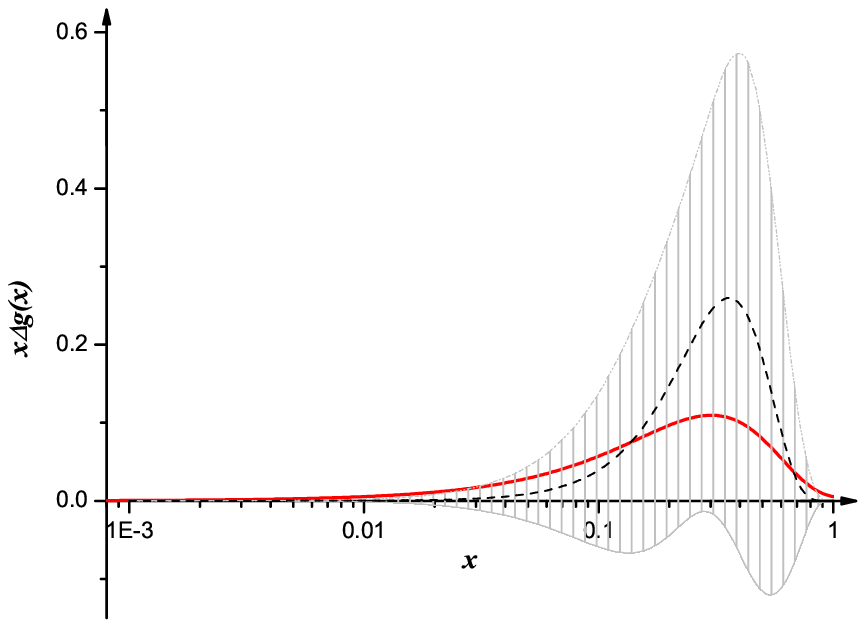}
\end{center}
\caption{$\Delta g(x)$ and $x\Delta g(x)$ calculated in the MIT bag
model. On the left panel, the results show successive additions of
$s$, $p$, $d$, $f$ wave contributions. In the second panel, the
result (red solid line) is compared with that from phenomenological
fit (dashed line surrounded by an uncertainty band).}
\label{Feynman-dg2}
\end{figure}

The phenomenological gluon distributions have been obtained by
fitting the $Q^2$ evolution of the spin-dependent structure function
$g_1(x)$ \cite{fits}. The result depends on the functional form
assumed for $\Delta g(x)$. In a recent study, the double spin
asymmetries for pion production was also included in the fits
\cite{hirai}. If one allows $\Delta g(x)$ to change sign, the fit
generates a distribution with very large error bars. On the other hand, if
one assumes that $\Delta g(x)$ is positive everywhere, the error
becomes much smaller. On the right panel of Fig. 2, we have shown
such a fit (dashed line with error band) together with our bag model
result (solid line). The bag calculation is consistent with
the fit, with significant strength at large and small $x$ where the model
might not be trustable.

To see that the positive $\Delta g(x)$ is a generic feature of quark
models, we have shown in Fig. 3 the result from a non-relativistic
quark model. The different curves again show successive inclusion
of higher intermediate excited states. The general shape is similar to
that from the MIT bag, which is shown in the thin solid line.
The model artifacts at small and large $x$ are clearly stronger.

\begin{figure}[hbt]
\begin{center}
\includegraphics[width=7cm]{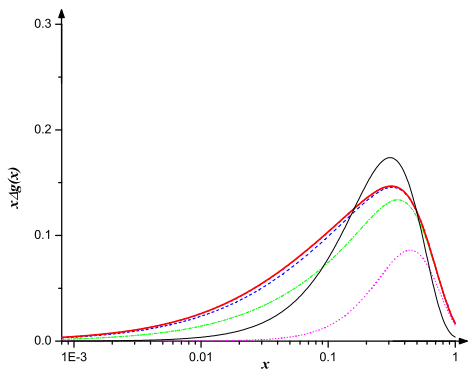}
\end{center}
\caption{$x\Delta g(x)$ calculated in non-relativistic quark model
by summing contributions from $s$, $p$, $d$, and $f$ waves. The thin
solid line is the MIT bag model result.} \label{Feynman-dg2}
\end{figure}

Integrating over $x$, the $\kappa = -1$ intermediate state produces
a result $\Delta G = 0.23\hbar $ (with $\alpha_s = 2.55$ obtained by
fitting $N-\Delta$ mass splitting). Including higher states up to
$\kappa = 3, -4$, we find $\Delta G = 0.32\hbar $, from which
a smaller $\alpha_s$ at higher excitation energy is used.
Therefore, at low-energy scales, $\Delta G$ is on the order of
0.2 to 0.3$\hbar$, which is considerably smaller than previous
expectations \cite{ar}. Indeed, the anomaly contribution from this $\Delta
G$
is negligibly small, $(\alpha_s/2\pi)\Delta G \sim 0.01$.

To summarize, we have argued that the polarized gluon distribution
$\Delta g(x)$ is calculable in quark models. We have carried out a
first such calculation in the MIT bag, and the result shows that it is
positive definite at all $x$. The total gluon helicity in the bag is
on the order of $0.2-0.3\hbar$, which is substantially smaller than
what has been expected in the past.

We are supported by the U. S. Department of Energy via grant
DE-FG02-93ER-40762. X. J. is also supported by a grant from NSFC.

\end{document}